\documentclass[journal]{IEEEtran}
\ifCLASSINFOpdf
\else
\fi
\usepackage{amsmath}
\usepackage{graphicx}
\usepackage{color}
\usepackage{placeins}
\usepackage{float}
\usepackage{hyperref}
\usepackage{tabularx,colortbl}
\usepackage{cuted}

\usepackage{slashbox}

\begin{document}
%
\title{The Effect of Inhomogeneous Dielectric Loading on Transmission Through a Slot in
an Infinite Thick Metallic Shield}
%
%
%

\author{Abdulaziz~H.~Haddab,~
        and~Edward~F.~Kuester,~
\thanks{The authors are with the Department of Electrical, Computer and Energy Engineering, University of Colorado, Boulder, CO 80309 USA e-mail: (haddab@Colorado.edu, Edward.Kuester@Colorado.edu)}
}

\maketitle

\begin{abstract}
Transmission through a slot in a an infinite metallic shield of finite thickness, inhomogeneously loaded with dielectric material, is modeled using an analytical approximation based on the slot being small compared to a free space wavelength. Resonances of relatively broad bandwidth (Fabry-Perot) as well as of very narrow bandwidth (Fano) are found. We show that by introducing a gap of different dielectric constant within the slot, we can control the separation between the broad bandwidth resonances, and the locations and magnitudes of the very narrow bandwidth resonances. The influence of grooves on both kinds of resonances is considered using the same analytical model. The effect of losses in the dielectric slot is also considered, using data for commercially available dielectrics. 
\end{abstract}


%
\IEEEpeerreviewmaketitle

\section{Introduction}
Transmission through a perforated metallic screen where the hole is much smaller than or comparable to the wavelength of the incident field has been a subject of intensive study. In particular, it has been shown that strong transmission can occur through a single slot \cite{had1,nee,lit,auck,auck1,eom,taka,eom1} or an array of slots \cite{eom1,solo,kim,ebb,kue,med,del,park} under certain conditions. Many techniques have been used to solve for the transmission through single or multiple slots in a thick metallic screen: a Green's function formulation \cite{nee}, an integral equation with a mode-series expansion \cite{had1,lit,eom,kim}, an equivalent circuit admittance model \cite{auck,auck1}, a surface
impedance model \cite{del}, and a circuit theory model \cite{med,park}, to mention only a representative sample of work. In the articles cited above, there are two main mechanisms responsible for the strong transmission---a Fabry-Perot resonance \cite{taka,med,park} that has a relatively broad bandwidth and is mainly due to resonance within the slot, or what has been called extraordinary transmission \cite{med,park}, which is due to coupling among the slots in an array. The latter resonances are usually of very narrow bandwidth and have the characteristic asymmetric shape of a Fano resonance. Our previous work on a single dielectric-loaded slot in a thick metallic screen \cite{had,had2} showed that in addition to previously known resonant (Fabry-Perot) transmission based on the fundamental waveguide mode of the slot, there exist extremely narrowband (Fano) resonances associated with the presence of higher-order modes in the slot, rather than coupling to adjacent slots. Due to the shape and narrow bandwidth of these resonances, the authors referred to them as extraordinary transmission---metaphorically, since we are dealing with a single slot and extraordinary transmission is generally understood to be an array phenomenon.

In the above-mentioned articles, only a homogeneously loaded slot was considered. In some articles, the problem of an inhomogeneously shaped slot, homogeneously loaded \cite{har1} or inhomogeneously loaded \cite{har,vol,ged} has been formulated for purely numerical solution, but only in \cite{ged} were results presented for such a case, the inhomogeneity being an absorbing material so that any resonances were significantly damped. In the present study, we extend our earlier analysis to include the effect of introducing a gap region of different low-loss dielectric permittivity within the slot on resonances due to both the fundamental parallel-plate TEM mode and to a higher-order mode in the slot \footnote{A brief preliminary version of the work presented in this paper has been given in \cite{had3}.
}. Our study reveals how the thickness of the gap within the slot, as well
as the dielectric constant of the substance that fills the gap, can control the location and magnitude of resonances. We show that a specified resonance may be targeted and moved to another frequency without significantly affecting neighboring resonances. This approach opens a new degree of freedom in design that to the best of our knowledge has not been recognized before. Our results are validated by comparison with a full-wave numerical finite element simulation (ANSYS-HFSS) \cite{hfss}.
\section{Derivation of the fields}
\subsection{Transmission Factor of $TF$}
\begin{figure}[!t]
\centering
\scalebox{0.2}
{\includegraphics{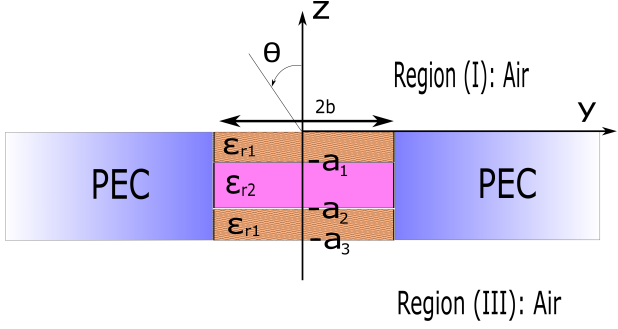}}
\caption{A metallic shield with an inhomogeneous slot.}
\label{Fig1Label}
\end{figure}
Similar to what we have done in previous work \cite{had,had2,had3}, let an H-polarized electromagnetic wave be obliquely incident at an angle $\theta$ to a single inhomogeneous dielectric-loaded slot of width $2b$ in a perfectly conducting metallic shield of finite thickness $a_3$ as shown in Figure~\ref{Fig1Label}. The slot consists of three layers of different thicknesses, where the 1st and 3rd layers of thickness $a_1$ and $a_3-a_2$ respectively are filled by a material of relative permittivity $\varepsilon_{r1}$, and the layer in the middle of thickness $a_2-a_1$ is filled by a material with $\varepsilon_{r2}$. We assume for simplicity that the gap is symmetrically located ($a_3-a_2=a_1$).

The formulation is based on an integral representation of the field external to the screen, and a mode-series expansion within the slot, as used in \cite{lit,had,had2,had3}. The magnetic field is expressed as: 
\begin{eqnarray}
\lefteqn{H_x(y,z) = }\nonumber \\
&& e^{-ik(\alpha y-\gamma z)}+e^{-ik(\alpha y+\gamma z)}+\int\limits_{-\infty}^{\infty} a(\xi)e^{-ik(\sqrt{1-\xi^2}z+\xi y)}\,d\xi, \nonumber \\
&& \qquad \qquad \qquad \qquad \qquad \qquad \qquad \qquad \qquad \qquad \mbox{($z>0$)} \nonumber \\
&& \sum\limits_{m=0}^\infty [A_m e^{-ih_m(z+a_1)}+B_m e^{ih_mz}] \cos{ \frac{\pi m}{2b}(y+b) }, \nonumber \\ 
&& \qquad \qquad \qquad \qquad \qquad \qquad \qquad \qquad \quad \mbox{($-a_1<z<0$)} \nonumber \\
&& \sum\limits_{n=0}^\infty [C_n e^{-ig_n(z+a_2)}+D_n e^{ig_n(z+a_1)}] \cos{ \frac{\pi n}{2b}(y+b) }, \nonumber \\
&& \qquad \qquad \qquad \qquad \qquad \qquad \qquad \qquad \mbox{($-a_2<z<-a_1$)} \nonumber \\
&& \sum\limits_{m=0}^\infty [E_m e^{-ih_m(z+a_3)}+F_m e^{ih_m(z+a_2)}] \cos{ \frac{\pi m}{2b}(y+b) }, \nonumber \\
&& \qquad \qquad \qquad \qquad \qquad \qquad \qquad \qquad \mbox{($-a_3<z<-a_2$)} \nonumber \\
&& \int\limits_{-\infty}^{\infty} d(\xi)e^{ik(\sqrt{1-\xi^2}(z+a_3)-\xi y)}\,d\xi, \quad \mbox{ ($z<-a_3$)} \label{eq1}
\end{eqnarray}
where $h_m=\sqrt{k^2\varepsilon_{r1}-(\frac{\pi m}{2b})^2}$, $g_n=\sqrt{k^2\varepsilon_{r2}-(\frac{\pi n}{2b})^2}$, $\alpha=\sin\theta$, $\gamma=\cos\theta$ and $A_m$, $B_m$, $C_n$, $D_n$, $E_m$ and $F_m$ are mode amplitudes. Applying the boundary condition of continuity of $H_x$ and $\frac{1}{\tilde{k}^2}\frac{\partial H_x}{\partial z}$ (where $\tilde{k}=k\sqrt{\varepsilon_r}$) at the interfaces between the dielectrics within the slot ($|y|< b$) at $z=-a_1$ and $z=-a_2$, we can obtain the following relations between $A_m$, $C_n$, $D_n$, $F_m$, $B_m$ and $E_m$:
\[
 A_m=\frac{e^{-ih_m a_1}}{T_{1m}+T_{2m}}\left[E_m- T_{3m} B_m \right]
\]
\[
 F_m=B_m e^{-ih_m a_1}\left[T_{1m}-T_{2m}+\frac{T_{3m}^2}{T_{1m}+T_{2m}}\right]
\]
\[
-\frac{T_{3m}e^{-ih_m a_1}}{T_{1m}+T_{2m}}E_m
\]
where
\[
 T_{1m}=\cos g_m(a_2-a_1), \quad T_{2m}=\frac{i(t_m^2 + 1)}{2t_m} \sin g_m(a_2-a_1)
\]
\[
 T_{3m}=\frac{i(t_m^2 - 1)}{2t_m} \sin g_m(a_2-a_1) \quad \mbox{\rm and} \quad t_m=\frac{g_m \varepsilon_{r1}}{h_m\varepsilon_{r2}}
\]
Mode orthogonality shows that modes with $m \neq n$ do not couple with each other. We next apply the boundary condition $\frac{\partial H_x}{\partial z}=0$ at the top and bottom surfaces of the metallic shield ($z=0$, $z=-a_3$ and $|y|> b$) and continuity of $H_x$ and $\frac{1}{\tilde{k}}\frac{\partial H_x}{\partial z}$ at the slot ($z=0$, $z=-a_3$ and $|y|< b$). Using the Fourier transform and defining $\tilde{x}^\pm(\zeta)=[a(\zeta)\pm d(\zeta)]\sqrt{1-\zeta^2}$ (where a new variable $\zeta$ has been introduced to avoid confusion with the variable of integration $\xi$), we arrive at the integral equations:
\begin{align}
& \tilde{x}^\pm(\zeta) = \frac{4\alpha(kb)^2\zeta}{\pi \varepsilon}  \sum\limits_{m=0}^{\infty} \frac{h_m b\Gamma_m^\pm}{(1+\delta_{0m})} G_m(\zeta) G_m(\alpha) \nonumber \\
 & +\frac{2(kb)^2\zeta}{\pi \varepsilon}  \sum\limits_{m=0}^{\infty} \frac{h_m b\Gamma_m^\pm}{(1+\delta_{0m})} \int\limits_{-\infty}^{\infty} \tilde{x}^\pm(\xi)\frac{\xi}{\sqrt{1-\xi^2}} G_m(\zeta) G_m(\xi) \,d\xi \label{x_sum}
\end{align}
where
$$G_m(\zeta) =  \frac{\sin(kb\zeta-\frac{\pi m}{2})}{(kb\zeta)^2-(\frac{\pi m}{2})^2}, \quad \Gamma_m^\pm=\frac{\Gamma_{2m}^\pm}{\Gamma_{1m}^\pm}$$
$$\Gamma_{1m}^\pm=\frac{e^{-i2h_m a_1}}{T_{1m}+T_{2m}} \pm 1 \mp \frac{T_{3m} e^{-i2h_m a_1}}{T_{1m}+T_{2m}}$$ 
$$\Gamma_{2m}^\pm=\frac{e^{-i2h_m a_1}}{T_{1m}+T_{2m}} \mp 1 \mp \frac{T_{3m} e^{-i2h_m a_1}}{T_{1m}+T_{2m}}$$ and $\delta_{0m}$ is the Kronecker delta.

For simplicity, we now assume normal incidence ($\theta=0$), and further that $kb\sqrt{\varepsilon_{r1}}<2\pi$; the latter condition ensures that only the $m=$ 0, 1 and 2 modes of the parallel plate waveguide will be above cutoff and all other modes will be evanescent (and for normal incidence, the $m=1$ mode is not excited). Retaining only these modes in (\ref{x_sum}), we get a degenerate kernel integral equation whose solution can be found by well-known methods \cite{had2}:
\begin{equation}\label{x}\tilde{x}^\pm(0)=\frac{2kb}{\pi \sqrt{\varepsilon_{r1}}} \left[\frac{N_0^\pm}{1+\frac{2 h_0 h_2 b^2  q^2 N_0^\pm N_2^\pm}{\pi^6 \varepsilon_{r1}^2} }\right]\end{equation} 
where $N_0^\pm=\frac{\Gamma_0^\pm}{1-\frac{kb\Gamma_0^\pm}{2\sqrt{\varepsilon_{r1}}} I_{00}}$ and
$N_2^\pm=\frac{\Gamma_2^\pm}{1-\frac{h_2 b \Gamma_2^\pm}{\varepsilon_{r1}}I_{22}}$, while for  $kb \ll 1$ \cite{lit}:
$$I_{00}\simeq 2+\frac{4i}{\pi}\ln\left(\frac{\phi}{kb}\right)-\frac{(kb)^2}{3}\left[1-\frac{i2}{\pi}\left(\ln(kb)-\frac{19}{12}+\gamma_E\right)\right]$$
and $I_{22}\simeq\frac{2i}{\pi^2} Si(2\pi)\simeq i0.2874\ldots$, $Si$ being the sine integral function \cite{had}.
Here, $\ln\phi=\frac{3}{2}-\gamma_E$ with $\gamma_E\simeq 0.5772 \ldots $ being Euler's constant.

Now we can determine the magnitude of the transmission through the infinite dielectric slot. The following integral will be defined as the (amplitude) transmission factor ($TF$):
\begin{eqnarray}
TF &=& \frac{\int\limits_{-b}^{b} E_y(y,-a_3) \,dy}{\int\limits_{-b}^{b}|E^{inc}(y,0)|\,dy} \label{TF} \\
&=& \frac{1}{\sqrt{\varepsilon_{r1}}}\left[\frac{N_0^+}{1+\frac{2 h_0 h_2 b^2  q^2 N_0^+ N_2^+}{\pi^6 \varepsilon_{r1}^2} }-\frac{N_0^-}{1+\frac{2 h_0 h_2 b^2  q^2 N_0^- N_2^-}{\pi^6 \varepsilon_{r1}^2} }\right] \nonumber
\end{eqnarray}
having used (\ref{x}) to obtain the final formula.

\subsection{Transmission Factor fundamental mode $TF_0$}
Analogous to what was done in \cite{had}, in this section we will present a simplified result that takes only the TEM mode into account. We then get a transmission factor that only contains the effect of the fundamental mode $m=0$:
\begin{equation}\label{TE0}TF_0=\frac{1}{\sqrt{\varepsilon_{r1}}} (N_0^+ - N_0^-)\end{equation} 
The reader may notice the similarity between the formulas for the homogeneous case \cite{had,had2} and those for the inhomogeneous case presented in this article; the only difference is in the definition of the $\Gamma_m^\pm$ parameters in each case. 

\section{Results}
Our formula allows us to examine two different problems or geometries. When we have $\varepsilon_{r1}>\varepsilon_{r2}$ we have what we call the gap geometry; on the other hand, when $\varepsilon_{r1}<\varepsilon_{r2}$ we have what we call a groove. The homogeneously filled slot showed two kinds of resonances: broad resonances due to the TEM mode, and sharp (Fano) resonances caused by the higher-order mode ($m=2$) \cite{had,had2}. Below, we investigate the effect of a gap or a groove on each type of resonance. A full description of the simulation setup for ANSYS HFSS can be found in \cite{had}.

\subsection{Gap Geometry}

\subsubsection{Fundamental-mode formula ($TF_0$)}
\begin{figure}[!t]
\centering
\scalebox{0.6}{\includegraphics{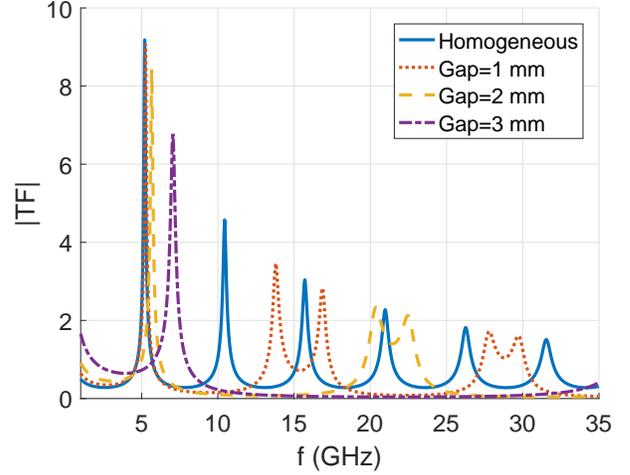}}
  \caption{Transmission Factor $TF_0$ for slot with $a_3=4$ mm, $b=1$ mm and different gap size= 0, 1, 2 and 3 mm}\label{Fig2}
\end{figure}

\begin{figure}[!t]
\centering
\scalebox{0.6}{\includegraphics{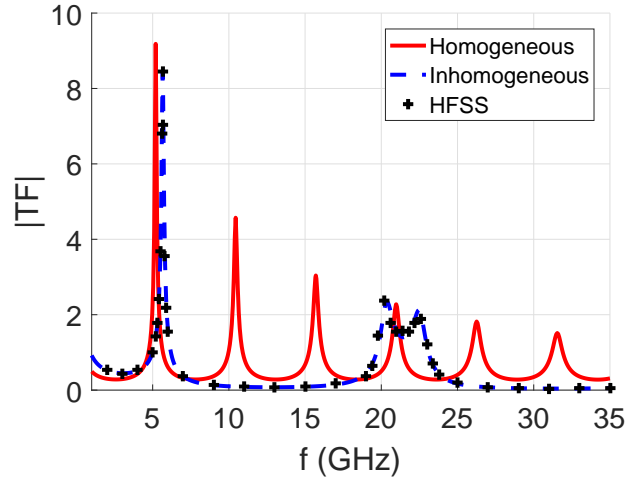}}
  \caption{Transmission Factor $TF_0$ of homogeneous and inhomogeneous slots: $a_3=4$ mm, $b=1$ mm and gap = 2 mm}\label{Fig3}
\end{figure}

\begin{figure}[!t]
\centering
\scalebox{0.3}{\includegraphics{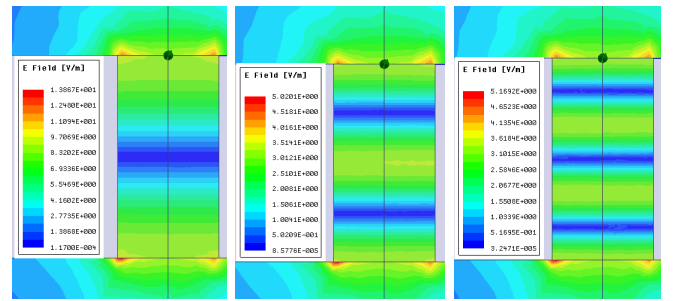}}
  \caption{Electric field distribution within the slot at first (odd, f=5.2 GHz), second (even, f=10.4 GHz) and third (odd, f=15.7 GHz) resonances for the homogeneous case $a_3=4$ mm, $b=1$ mm (no gap).}\label{field}
\end{figure}

\begin{figure}[!t]
\centering
\scalebox{0.25}{\includegraphics{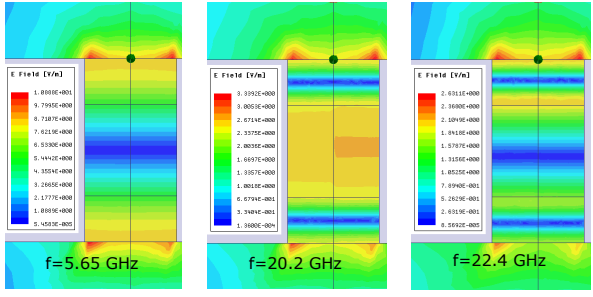}}
  \caption{Electric field distribution within the slot at first (odd, f=5.65 GHz), second (even, f=20.2 GHz), third (odd, f=22.4 GHz) resonances for the inhomogeneous case $a_3=4$ mm, $b=1$ mm and gap= 2 mm, reference figure~\ref{Fig3}.}\label{field1}
\end{figure}

We assume a shield thickness $a_3=4$ mm and a slot width $2b=2$ mm, with $\varepsilon_{r1}=50$, $\varepsilon_{r2}=1$ and a symmetrically located gap: $a_3-a_2=a_1$. A comparison for different gap sizes (no gap, $a_2 - a_1 = 1$ mm , 2 mm and 3 mm) determined from equation (\ref{TE0}) is shown in Figure~\ref{Fig2}. A comparison between the result of equation (\ref{TE0}), the analytical result for the case of a homogeneous slot (no gap) \cite{had}, and a full-wave finite-element (FEM) simulation using ANSYS HFSS for gap size $(a_2-a_1)=2$ mm is shown in Figure~\ref{Fig3}.
The results show that introducing a gap within the slot will shift the even-order resonances to higher frequencies while keeping odd-order resonances almost unchanged.  Increasing the size of the gap will shift the even-order resonances to progressively higher frequencies until they meet the next odd-order resonances, which begin to be shifted to higher frequencies but at a rate much slower than for the even-order resonances, eventually forming new resonances which keep moving to higher frequencies as the size of the gap is increased. Because the gap is introduced in the middle of the slot in our example, the electric field of the odd-order resonances is minimum at that location, but is at a maximum for even-order resonances as can be seen in Figure~\ref{field} for homogeneous and in Figure~\ref{field1} for inhomogeneous, which explains why the impact of the gap on the latter is larger.

\subsubsection{Resonances from higher order modes ($TF$)} 
In the previous section (also see \cite{had3}), we saw that the gap size controls the frequency intervals between TEM resonances. In \cite{had} and \cite{had2}, we showed that including higher order modes in the analysis of the homogeneous slot reveals new sharp resonances that are always located above the cutoff frequency of the second parallel-plate waveguide mode ($m=2$). We obtained an analytical formula to locate these new resonances in the frequency domain.

\begin{figure}[!t]
\centering
\scalebox{0.6}{\includegraphics{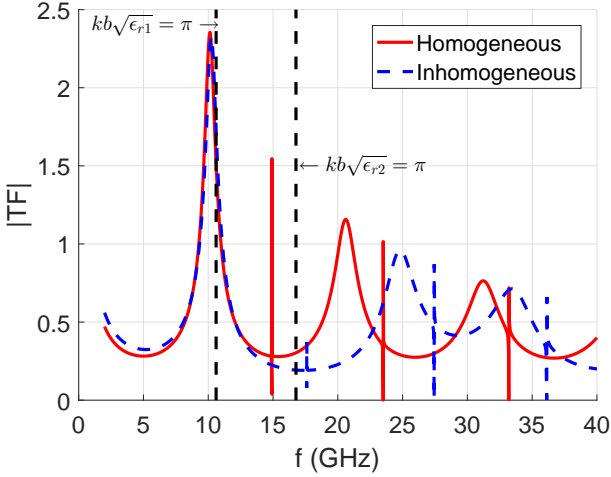}}
  \caption{Transmission Factor $TF$ of homogeneous and inhomogeneous symmetrically filled ($a_3-a_2=a_1$) slot: $a_3=2$ mm, $b=2$ mm, $\varepsilon_{r1}=50$, $\varepsilon_{r2}=20$ and gap = 0.6 mm.}\label{Fig4}
\end{figure}

\begin{figure}[!t]
\centering
\scalebox{0.6}{\includegraphics{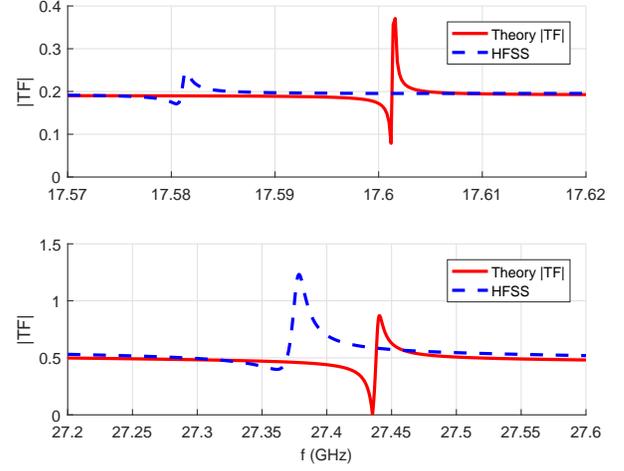}}
  \caption{Analytical transmission Factor $TF$ of inhomogeneous and numerical results (HFSS)  near 17.6 GHz for symmetrically filled  ($a_3-a_2=a_1$) slots: $a_3=2$ mm, $b=2$ mm, $\varepsilon_{r1}=50$, $\varepsilon_{r2}=20$ and gap = 0.6 mm.}\label{Fig5}
\end{figure}

In this section, we will investigate the effect of introducing a gap within the slot on higher order mode resonances. For parameter values $a_3=2$ mm, $2b=4$ mm, $a_2=1.3$ mm, $\varepsilon_{r1}=50$ and $\varepsilon_{r2}=20$, a comparison between results of the inhomogeneous gap formula (\ref{TF}), the homogeneous slot (no gap) formula from \cite{had} and \cite{had2} is shown in Figure~\ref{Fig4}. Also, comparisons between results of the inhomogeneous gap formula (\ref{TF}) and a full-wave ANSYS HFSS simulation near 17.6 GHz and 27.4 GHz are shown in Figure~\ref{Fig5}.
The numerical results show excellent agreement with the analytical formula for the frequency location of the Fano resonances. There is some disagreement between the amplitudes at the Fano resonant frequency, but this is typical for full-wave simulations of resonances with such narrow bandwidth \cite{had}.

The results show that the higher order resonances' locations and magnitudes will be determined by the gap's size and its dielectric permittivity $\varepsilon_{r2}$. Increasing the size of the gap will always shift the higher order mode resonances to higher frequencies. The choice of dielectric constant will determine the cut-off frequency of the higher-order parallel-plate waveguide mode within the gap (in our case, the second mode $m=2$) which represents the threshold between propagation and evanescence of the mode within the gap. So if we choose a dielectric such that the sharp resonance is located below that threshold (cut-off frequency), then increasing the gap size will shift the resonance to a higher frequency. If it doesn't reach and pass the threshold, it will be attenuated and vanish, but if it passes this threshold, it will be sustained and keep shifting to higher frequencies as the size of the gap is increased.

\begin{figure}[!t]
\centering
\scalebox{0.35}{\includegraphics{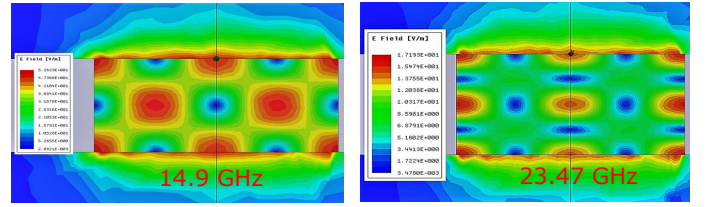}}
  \caption{Electric field distribution (magnitude) within the slot for higher order resonances of homogeneous case near 15 and 23.5 GHz (reference \ref{Fig4}), $a_3=2$ mm, $b=2$ mm, $\varepsilon_{r}=50$.}\label{2nd-dis}
\end{figure}

\begin{figure}[!t]
\centering
\scalebox{0.2}{\includegraphics{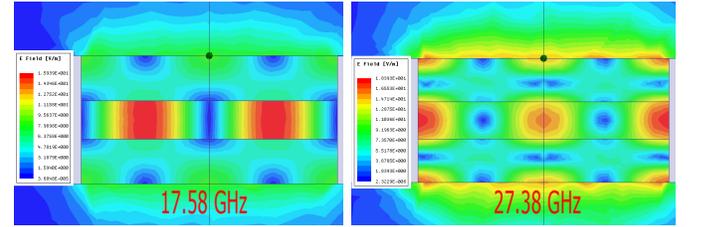}}
  \caption{Electric field distribution (magnitude) within the slot for higher order resonances of inhomogeneous case near 17.58 and 27.38 GHz (reference \ref{Fig4}), $a_3=2$ mm, $b=2$ mm, $\varepsilon_{r1}=50$, $\varepsilon_{r2}=20$ and gap=0.6 mm.}\label{2nd-dis-2}
\end{figure}

\begin{figure}[!t]
\centering
\scalebox{0.45}{\includegraphics{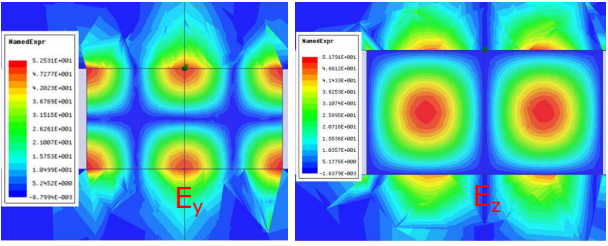}}
  \caption{Electric field distribution (y and z components) within the slot for higher order resonances of homogeneous case near 15 GHz (reference \ref{2nd-dis}), $a_3=2$ mm, $b=2$ mm, $\varepsilon_{r}=50$.}\label{y-z-1}
\end{figure}

\begin{figure}[!t]
\centering
\scalebox{0.5}{\includegraphics{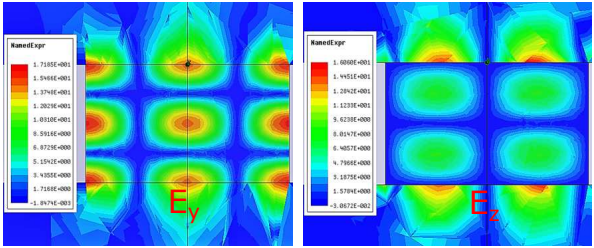}}
  \caption{Electric field distribution (y and z components) within the slot for higher order resonances of homogeneous case near 23.5 GHz (reference \ref{2nd-dis}), $a_3=2$ mm, $b=2$ mm, $\varepsilon_{r}=50$.}\label{y-z-2}
\end{figure}

The electric field distribution of those resonances in both cases homogeneous and inhomogeneous cases are shown in figures~\ref{2nd-dis} and \ref{2nd-dis-2}. 
It is obvious that the electric field distribution of the first higher order resonance is disturbed by introducing the gap while the second one is almost unchanged. The electric field components ($E_y$ and $E_z$) for the first and second higher order resonances for homogeneous case are shown in Figures~\ref{y-z-1} and \ref{y-z-2} respectively. The (tangential) $E_y$ component will be continuous while the normal one ($E_z$) is not, so introducing the gap in the middle disturbs $E_z$ in the higher order resonance near 15 GHz where the field is maximum, but has almost no effect near 23.5 GHz where the field is minimum. 

To see whether realistic materials with loss will alter the foregoing results, we consider the following commercially available dielectric, for which $\varepsilon_{r1}=75$ and the loss tangent is $\tan\delta =2.6\times 10^{-4}$ at 3 GHz and $\varepsilon_{r2}=35$ and the loss tangent is $\tan\delta =2.9\times 10^{-5}$ at 2 GHz \cite{skywork}. For a fair comparison, we move the resonant frequency to around 3 GHz by modifying the slot size to $a_3=8$ mm and $b=16$ mm.
\begin{figure}[!t]
\centering
\scalebox{0.6}{\includegraphics{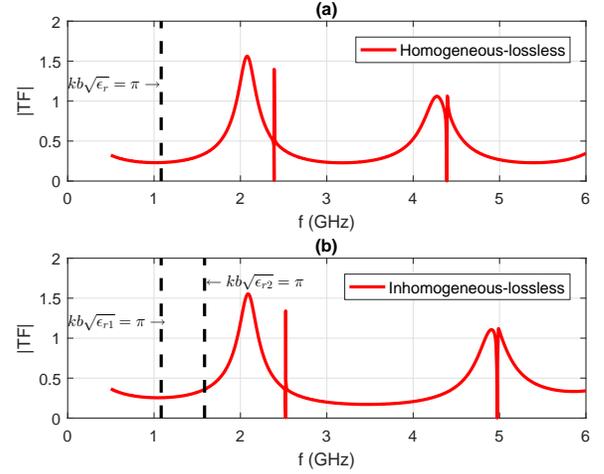}}
  \caption{Transmission Factor $TF$ lossless case, $a_3=8$ mm, $b=16$ mm: (a) Homogeneous: $\varepsilon_{r}=75$. (b) Inhomogeneous: $\varepsilon_{r1}=75$, $\varepsilon_{r2}=35 $ and gap=2 mm}\label{Fig7}
\end{figure}
\begin{figure}[!t]
\centering
\scalebox{0.6}{\includegraphics{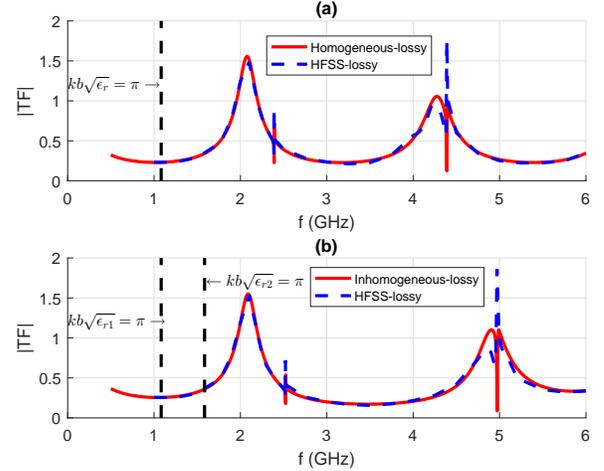}}
  \caption{Transmission Factor $TF$ and numercial results (HFSS), lossy case, $a_3=8$ mm, $b=16$ mm: (a) Homogeneous: $\varepsilon_{r}=75(1-i2.6\times 10^{-4})$. (b) Inhomogeneous: $\varepsilon_{r1}=75(1-i2.6\times 10^{-4})$, $\varepsilon_{r2}=35(1-i1\times 10^{-4})$ and gap=2 mm}\label{Fig8}
\end{figure}
Results are shown in Figure~\ref{Fig7} for the lossless case and in Figure~\ref{Fig8} for the lossy case and compared with numerical results from HFSS. We see that the broad TEM mode resonances are barely affected by losses within the dielectric slot, while the sharp resonances, though still present, are changed significantly in amplitude.

\subsection{Groove Geometry}

\subsubsection{Fundamental-mode formula ($TF_0$)}
\begin{figure}[!t]
\centering
\scalebox{0.6}{\includegraphics{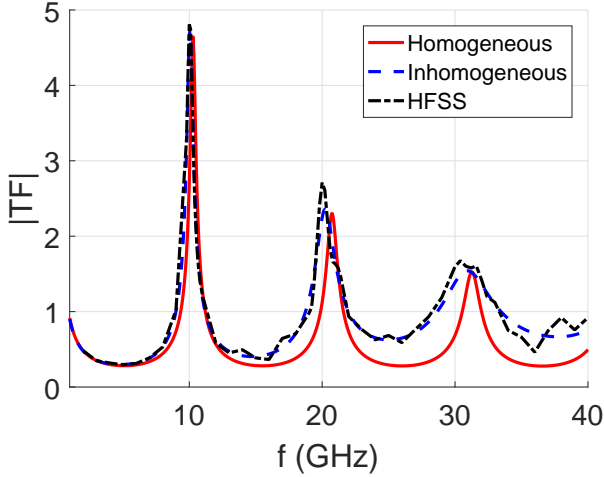}}
  \caption{Transmission Factor $TF_0$ of homogeneous ($a=2$ mm, $b=1$ mm and $\varepsilon_r=50$) and inhomogeneous ($a_3=4$ mm, $b=1$ mm, $a_1=1$mm (groove=1 mm), $\varepsilon_{r1}=1$ and $\varepsilon_{r2}=50$). Note: The thickness in homogeneous case is 2mm which equal the thickness of dielectric in inhomogeneous case}\label{Fig9}
\end{figure}

\begin{figure}[!t]
\centering
\scalebox{0.6}{\includegraphics{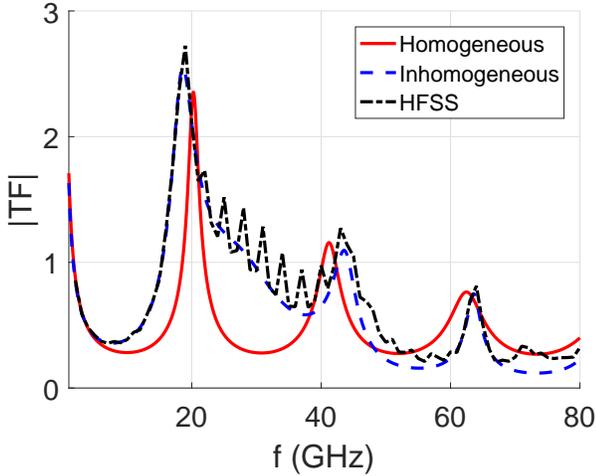}}
  \caption{Transmission Factor $TF_0$ of homogeneous ($a=1$ mm, $b=1$ mm and $\varepsilon_r=50$) and inhomogeneous ($a_3=4$ mm, $b=1$ mm, $a_1=1.5$mm (groove=1.5 mm), $\varepsilon_{r1}=1$ and $\varepsilon_{r2}=50$). Note: The thickness in homogeneous case is 1mm which equal the thickness of dielectric in inhomogeneous case}\label{Fig10}
\end{figure}

\begin{figure}[!t]
\centering
\scalebox{1}{\includegraphics{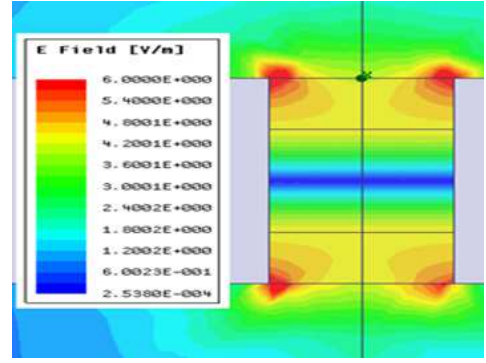}}
  \caption{Electric field distribution (magnitude) at 10 GHz for inhomogeneous case: $a_3=4$ mm, $b=1$ mm, $a_1=1$mm (groove=1 mm), $\varepsilon_{r1}=1$ and $\varepsilon_{r2}=50$.}\label{groove}
\end{figure}

We assume a shield thickness $a_3=4$ mm and a slot width $2b=2$ mm with $\varepsilon_{r1}=1$ (air) and $\varepsilon_{r2}=50$ (symmetrically located: $a_3-a_2=a_1$). With this choice, we actually form a groove in the slot, in which we will fix the thickness of the shield (PEC) while varying the thickness of the dielectric layer. A comparison among the result of equation (\ref{TE0}), the analytical result for the case of homogeneous (no groove) \cite{had}, and a full-wave finite-element simulation using Ansys HFSS for different groove sizes $a_1=1$ and $1.5$ mm (the thicknesses of dielectric layer ($a_2-a_1$) within the slot being 2 mm and 1 mm respectively) are shown in Figures~\ref{Fig9} and \ref{Fig10}, where we compare with the homogeneous case that has thickness equal to the thickness of dielectric layer in inhomogeneous case. The electric field magnitude distribution at 10 GHz for a groove size $a_1=1$ mm is shown in Figure~\ref{groove}.
The results show that introducing a groove in the slot has less effect on the first resonance, where it slightly changes its resonance frequency and magnitude. On the other hand, the groove significantly alters the magnitude of the Transmission Factor at the higher resonance frequencies and away from those resonances.

If we introduce the air groove but keep the screen thickness constant, we actually reduce the thickness of dielectric layer, and that will alter the ratio ($\frac{a}{b}$), which will change the magnitude of the broad resonances that can be determined using equation (19) in \cite{had}. We note in Figure~\ref{Fig10} that there appear many closely spaced "ripples" in the HFSS results between 20 and 40 GHz. By plotting the field distribution in HFSS, we observe standing waves at the peaks of the ripples in Figure~\ref{Fig10}. We believe that this artifact is due to imperfect absorption of the radiation boundary at those particular frequencies.

\subsubsection{Resonances from higher order modes ($TF$)}
\begin{figure}[!t]
\centering
\scalebox{0.6}{\includegraphics{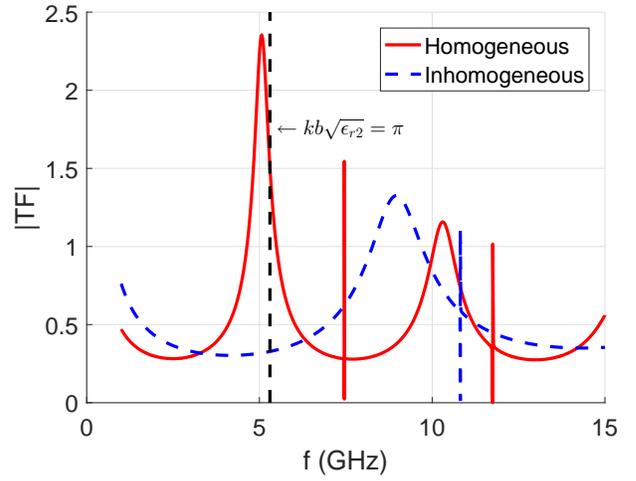}}
  \caption{Transmission Factor $TF$ of homogeneous and inhomogeneous symmetrically filled ($a_3-a_2=a_1$) slot: $a_3=4$ mm, $b=4$ mm, $\varepsilon_{r1}=1$, $\varepsilon_{r2}=50$ and groove= 0.9 mm.}\label{groove_2nd}
\end{figure}

\begin{figure}[!t]
\centering
\scalebox{0.6}{\includegraphics{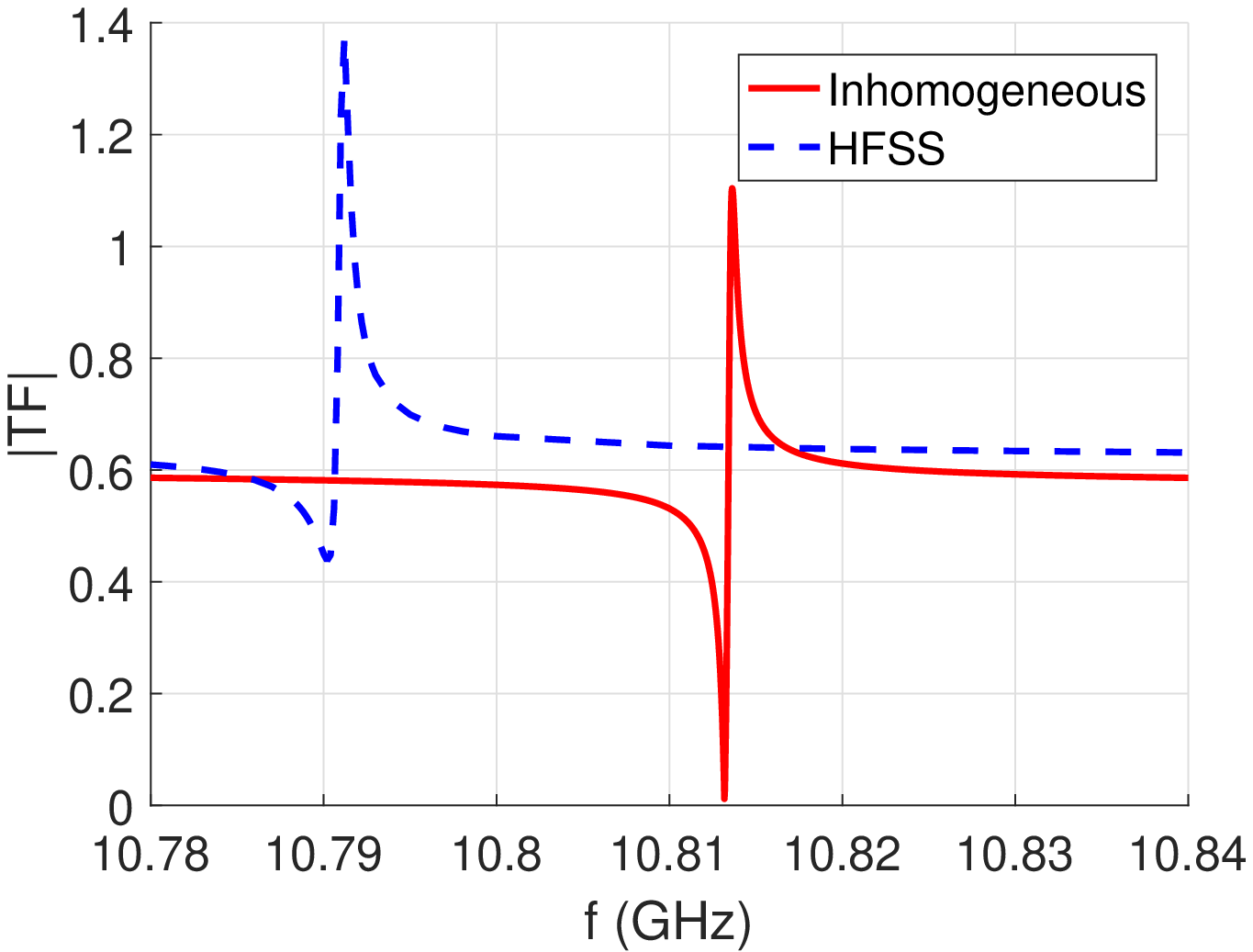}}
  \caption{Analytical transmission Factor $TF$ of inhomogeneous and numerical results (HFSS) near 10.8 GHz for sysmmetrically filled ($a_3-a_2=a_1$) slot: $a_3=4$ mm, $b=4$ mm, $\varepsilon_{r1}=1$, $\varepsilon_{r2}=50$ and groove= 0.9 mm.}\label{groove_2nd_hfss}
\end{figure}

In this subsection, we investigate the effect of a groove on the higher order mode resonances. Comparison between results of the homogeneous and inhomogeneous cases is shown in Figure~\ref{groove_2nd} for $a_3=4$ mm, $a_1=0.9$ mm (symmetrically located: $a_3-a_2=a_1$) and a slot width $2b=8$ mm with $\varepsilon_{r1}=1$ (air), $\varepsilon_{r2}=50$. 
As we expected, introducing the air grooves shifts all resonances to higher frequencies as the high dielectric layer decreases in size, with magnitude reduction in the broad resonances consistent with the prediction of equation~(19) and Table~I in \cite{had}. The comparison with HFSS results for the higher-order resonance at 10.81 GHz for the inhomogeneous case in Figure~\ref{groove_2nd} is shown in detail in Figure~\ref{groove_2nd_hfss}, and the agreement is seen to be good.

\section{Conclusion}
In this study, we have obtained an approximate analytical formula to determine the transmission through an inhomogeneously loaded dielectric slot in a thick metallic screen. This analytical model allows us to study two different loading geometries (a gap and grooves) and analyze the influence on both kinds of resonances (Fabry-Perot and Fano). By introducing a region of different dielectric constant within the slot (gap), we can control the separation between the broad bandwidth resonances (Fabry-Perot) and the locations and magnitudes of the very narrow bandwidth (Fano) resonances. In addition, the effect of creating grooves on both kind of resonances is considered, as well as the losses in the dielectric slot. Our results were validated by comparison with a full-wave numerical finite element simulation (HFSS).


%

\appendices


\ifCLASSOPTIONcaptionsoff
  \newpage
\fi



%

%

\vfill





\end{document}